\documentclass[aps,pra,twocolumn,superscriptaddress,amsmath,amssymb,showpacs]{revtex4-1}

\usepackage{graphicx}	
\usepackage{dcolumn}	
\usepackage{bm}			
\usepackage{verbatim} 	
\usepackage{braket}
\usepackage{xr}
\usepackage{hyperref}
\usepackage{enumerate}

\usepackage{color}
\usepackage{soul}
\usepackage{upgreek}

\begin{document}

\title{Percolation based architecture for cluster state creation using photon-mediated entanglement between atomic memories}

\author{Hyeongrak Choi}
\email{choihr@mit.edu}
\affiliation{Research Laboratory of Electronics, Massachusetts Institute of Technology, Cambridge, Massachusetts, 02139, United States}
\author{Mihir Pant}
\affiliation{Research Laboratory of Electronics, Massachusetts Institute of Technology, Cambridge, Massachusetts, 02139, United States}
\affiliation{Quantum Information Processing Group, Raytheon BBN Technologies, 10 Moulton Street, Cambridge, Massachusetts 02138, USA}
\author{Saikat Guha}
\affiliation{Quantum Information Processing Group, Raytheon BBN Technologies, 10 Moulton Street, Cambridge, Massachusetts 02138, USA}
\affiliation{College of Optical Sciences, University of Arizona, 1630 East University Boulevard, Tucson, AZ}
\author{Dirk Englund}
\email{englund@mit.edu}
\affiliation{Research Laboratory of Electronics, Massachusetts Institute of Technology, Cambridge, Massachusetts, 02139, United States}

\begin{abstract}
A central challenge for many quantum technologies concerns the generation of large entangled states of individually addressable quantum memories. Here, we show that percolation theory allows the rapid generation of arbitrarily large graph states by heralding the entanglement in a lattice of atomic memories with single-photon detection. This approach greatly reduces the time required to produce large cluster states for quantum information processing including universal one-way quantum computing. This reduction puts our architecture in an operational regime where demonstrated coupling, collection, detection efficiencies and coherence time are sufficient. The approach also dispenses the need for time consuming feed-forward, high-cooperativity interfaces and ancilla single photons, and can tolerate a high rate of site imperfections. We derive the minimum coherence time to scalably create large cluster states, as a function of photon-collection efficiency. We also propose a variant of the architecture with long-range connections, which is even more resilient to site yields. We analyze our architecture for nitrogen-vacancy (NV) centers in diamond, but the approach applies to any atomic or atom-like systems.
\end{abstract}

\maketitle

\setcounter{secnumdepth}{0}
\section{Introduction}
The past years have seen rapid advances in controlling small groups of qubits encoded in atomic or atom-like quantum memories. An important question now concerns the development of architectures to efficiently combine these memories into large-scale systems capable of general-purpose quantum computing~\cite{2014.PRX.Nemoto-Munro.DiamQCArch, 2014.PRA.Monroe-Kim.ModQC, 2014.PRX.Nickerson-Benjamin.LossyPhotonicLinkQC}, quantum simulation~\cite{2013.NatPhys.Cai-Plenio.DiamondQSim}, and metrology near the quantum limit~\cite{2011.NatPhot.Giovannetti-Maccone.QuantMetRev}.
A promising approach is entangling the atomic qubits with optical links to generate cluster states which can perform general-purpose quantum computing with adaptive single-qubit measurements~\cite{2001.PRL.Raussendorf-Briegel.ClusterStateComp}. A key challenge is to produce this cluster state fast enough to allow computation and error correction within a coherence time of memories. 

Here, we show that percolation of heralded entanglement allows us to create arbitrarily large cluster states. 
Our cluster state based architecture does not require reconfiguration of physical qubits by the result of the entanglement success unlike conventional approaches~\cite{duan2005efficient}. Instead, the optical network reconfigures the connectivity. Meanwhile, the concept of percolation greatly relaxes the high success probability required in the previous proposals  \cite{2014.PRX.Nemoto-Munro.DiamQCArch, 2014.PRA.Monroe-Kim.ModQC, 2014.PRX.Nickerson-Benjamin.LossyPhotonicLinkQC}. The process is fast enough for implementation with the device parameters demonstrated; one does not need high cooperativity cavities and ancilla single photons. It requires `feed-forward', operations conditioned on the previous measurement results. In the absence of errors, the missing bonds can be compensated with constant overhead~\cite{2007.PRL.Kieling-Eisert.PercolationQC, 2015.PRL.Gimeno-Segovia-Rudolph.3GHZtoBallisticQC, 2015.PRA.Zaidi-Rudolph.BallisticLOQC, 2008.NJP.Browne-Short.PercPhaseTrans}. In the presence of errors, the scheme can be adapted to be fault-tolerant \cite{2006.AnnPhys.Raussendorf-Goyal.FaultTolClus,2007.NJP.Raussendof-Goyal.TopoFTClus}. Our approach also provides tolerance for site imperfections, and we show trade-offs between reaching percolation threshold and site loss-resilience. Combined with transparent nodes implemented on nanophotonic platform, long range connections are possible, reducing the percolation threshold. We derive a theoretical lower bound on the minimum time required to percolate in any lattice, and found that the proposed lattices are only a factor of $1.6\sim3$ above this limit. 

We focus on nitrogen vacancy (NV) centers in diamond~\cite{2013.PhysRep.Doherty-Hollenberg.NVCentre} and propose the control sequence to map the physical properties to cluster state quantum computation. NV centers have favorable properties as quantum memories. The NV$^-$ charge state has a robust optical transition for heralded entanglement between remote NV centers~\cite{2013.Nature.Bernien-Hanson.3ment, 2015.Nature.Hensen-Hanson.BellTest} and a long electronic spin ($S$=1) coherence time approaching one second \cite{abobeih2018one}. Recently, single qubit gates with fidelities up to 99$\%$ were achieved with optimal control techniques \cite{2014.NatComm.Dolde-Wrachtrup.spinent}. Electronic spins of NV centers can be coupled with nearby nuclear spins, which have coherence times exceeding one minute \cite{bradley2019nuclear.one.minute}. In addition, they can be coupled with integrated nanophotonic devices \cite{2015.PRX.Mouradian-Englund.DiamWGonSiN}.

Our approach possibly applies to a number of physical systems, including atomic gases~\cite{2005.Nature.Chou-Kimble.measindent}, ion traps~\cite{2007.Nature.Moehring-Monroe.entbits}, semiconductor quantum dots~\cite{2015.NatPhys.Delteil-Imamoglu.HerEntDistSp}, and rare earth ions~\cite{2012.NatComm.Kolesov-Wrachtrup.detrareearthion}.

\section{Results}

\begin{figure}[t]
\includegraphics[width=\columnwidth, trim=10 10 10 10]{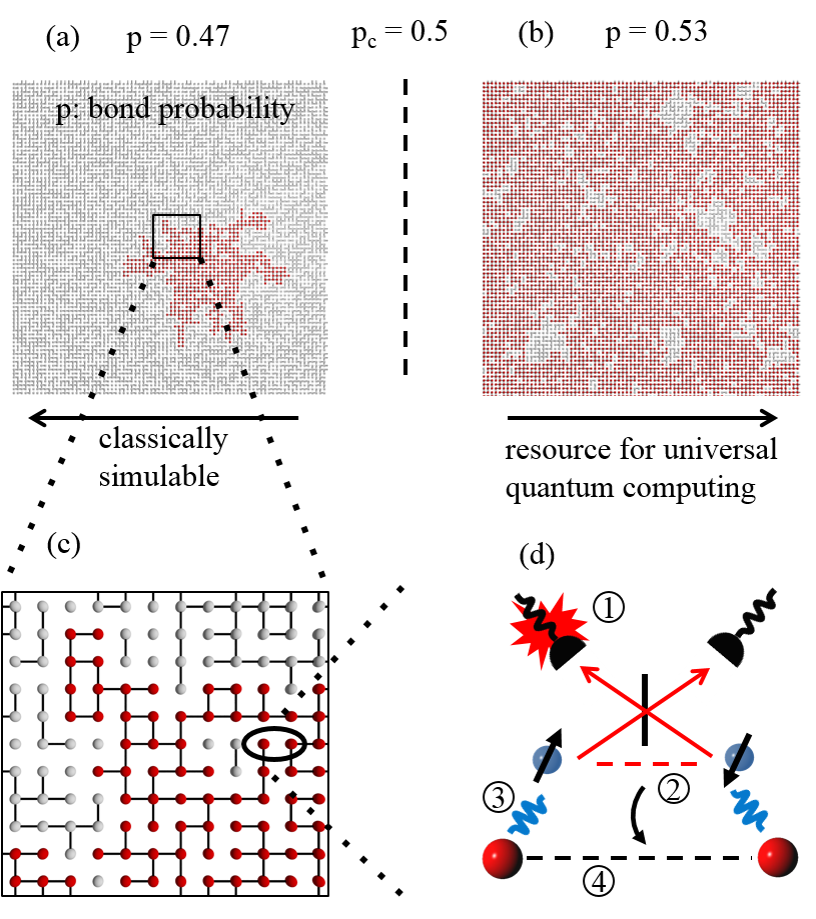}
\caption{Cluster state generation by percolation. (a),(b) Transition in the size of the largest connected component (LCC) with increasing bond probability. Spheres and lines represent nodes and bonds respectively, and the red spheres are in LCC. When the bond probability ($p$) goes above the percolation threshold ($p_{c}$), the size of the LCC suddenly increases. Corresponding graph states change from being classically simulable to a resource for universal quantum computation. (c) Expanded view of (a). (d) Physical implementation of nodes and bonds with NV centers in diamond. \textcircled{1} Probabilistic Bell measurement is attempted on two nearest-neighbor electronic spins (blue spheres). \textcircled{2} Conditioned on two single-photon detection events, the two electron spins are entangled (Bell states). \textcircled{3} Hyperfine interaction entangles electronic spins and nuclear spins ($^{15}$N) through controlled-Z gates. \textcircled{4} Measurement of electronic spins in transverse direction projects remote nuclear spins onto an entangled state (entanglement swapping).}
\centering
\label{fig:perc_transition}
\end{figure}

\subsection{Protocol}

Figure~\ref{fig:perc_transition} illustrates the percolation approach to generate cluster states with NV centers. We work in the framework of cluster states where nodes represent qubits in the superposition state $(\ket{0} + \ket{1})/\sqrt{2}$ and bonds represent the controlled-Z (CZ) gate. Consider a square lattice where edges exist with probability $p$ (Fig.~\ref{fig:perc_transition}(a)-(c)). The computational power of the cluster state corresponding to the graph is related to the size of the largest connected component (LCC) (shown in red). When $p < 0.5$, the graph produces small disconnected islands. For a lattice with $N$ nodes, the size of the LCC is $O(\log (N))$~\cite{2000.PRE.Bazant.SubCritPerc}. Single-qubit measurements on such a cluster state can be efficiently simulated on classical computers. When the bond probability $p$ exceeds the critical value $p_c=0.5$, called the percolation threshold of the lattice, there is a sudden transition in the size of the LCC. On crossing this boundary, the size of the LCC jumps to $\Theta(N)$ and the lattice is `percolated'. A large percolated lattice can be `renormalized' with single qubit measurements to obtain a perfect lattice \cite{2007.PRL.Kieling-Eisert.PercolationQC}. Renormalization consumes a constant fraction of qubits \cite{2007.PRL.Kieling-Eisert.PercolationQC} and requires classical computation (pathfinding) \cite{herr2018local,morley2017physical}. Resulting perfect lattice is a resource for universal quantum computation. Thus, the percolation transition is accompanied by a sudden transition in computational power; adaptive single-qubit measurements on the cluster state enable universal quantum computing~\cite{2008.NJP.Browne-Short.PercPhaseTrans}.

Figure~\ref{fig:perc_transition}(d) shows the physical implementation of bond creation (entanglement) with NVs. The nuclear spins (red spheres) are `client qubits' that store entanglement. They are coupled to NV electronic spins (`broker qubits'), that can be entangled remotely by photon-mediated Bell measurements. In each time step, we attempt to create one bond (entanglement) at each node. We consider the Barrett-Kok protocol ~\cite{2005.PRA.Barrett-Kok.entprot} on the broker qubits of neighboring nodes. If the probabilistic Bell measurement succeeds, the electron spins in each node are entangled. This entanglement is transferred to the nuclear spins with the entanglement swapping procedure, as illustrated in Fig.~\ref{fig:perc_transition}(d) \cite{supp_percatmem,2014.PRX.Nemoto-Munro.DiamQCArch}. The whole cycle from initialization to entanglement swapping is assumed to be approximately $t_0 = 5~\upmu$s based on experimental demonstrations~\cite{2015.Nature.Hensen-Hanson.BellTest, supp_percatmem}.

If the Bell measurement fails, one needs to restore the nuclear spins as before the trial. In addition, the electronic spins should be initialized to spin ground state without disrupting the nuclear spin. For the first problem, we just wait for the nuclear spin and electronic spin to be decoupled, which happens after a time period of the hyperfine interaction \cite{supp_percatmem}. This time period serves as a global clock and can be synchronized across the whole graph only using the $^{15}$N nuclear spin, which is the host atom of NV centers. For the electron spin initialization, optical pumping technique cannot be used as in many experiments~\cite{2013.Nature.Bernien-Hanson.3ment, 2015.Nature.Hensen-Hanson.BellTest}, because hyperfine coupling persists during a random amount of time when the electronic spin is in $\ket{m_s=1}$ states. Instead, we read out the spin state via $\ket{m_s=0}$ optical transition where the electronic spin decouples from the nuclear spin. If the spin is in the other state, one can use a fast microwave pulse to put it in the spin ground state \cite{supp_percatmem}.

We assume the Barrett-Kok protocol because it does not require ancilla single-photons or high cooperativity cavities, unlike reflection based protocols \cite{2014.PRX.Nemoto-Munro.DiamQCArch}. Furthermore, unlike single photon detection protocols \cite{cabrillo1999creation}, photon loss does not degrade the entanglement fidelity $\bra{\Psi}\hat{\rho}\ket{\Psi}$ where $\ket{\Psi}$ is the Bell state, and $\hat{\rho}$ is density matrix operator of the system after successful heralding. High entanglement fidelity is important for reducing the error correction overhead in a fault-tolerant architecture. High fidelity comes at the price of low bond success probability \cite{2005.PRA.Barrett-Kok.entprot}, that can be overcome in the percolation based architecture. 

\subsection{Physical unit}

\begin{figure}[t]
\includegraphics[width=\columnwidth]{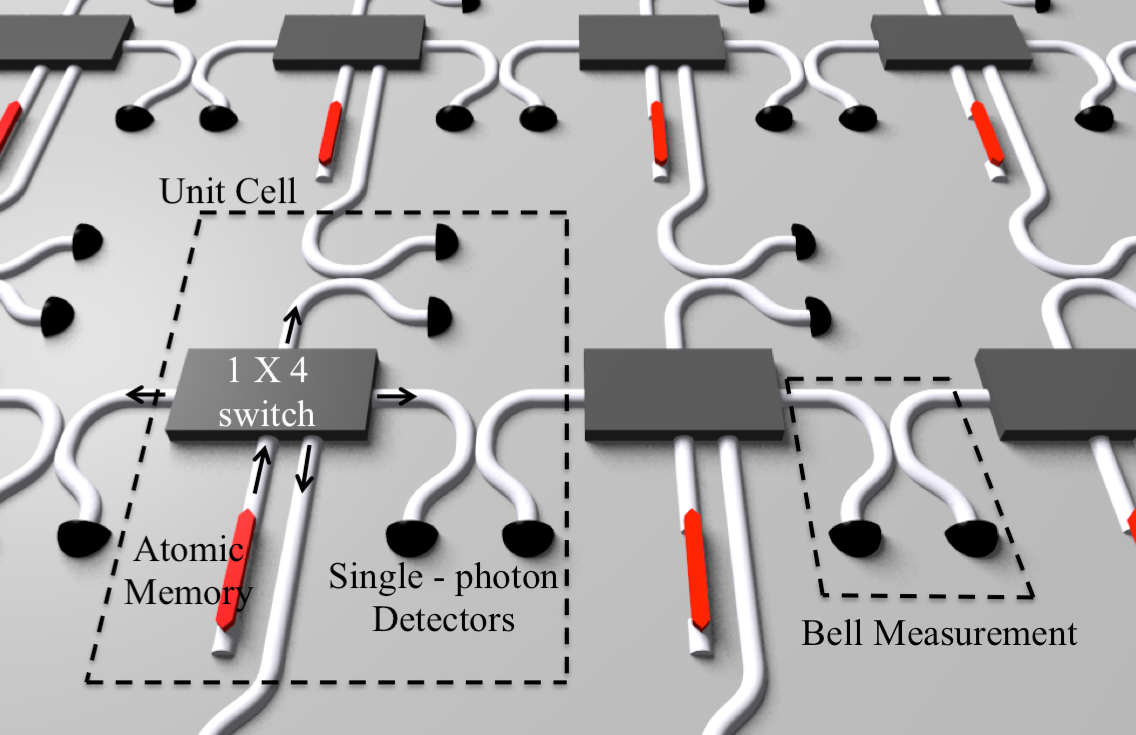}
\caption{Physical implementation of the proposed architecture. A unit cell consists of an atomic memory, a $1\times4$ switch, a 50/50 beam-splitter, waveguides and 4 single-photon detectors. Single-photons emitted from the atomic memory are coupled to the waveguide and directed to the switch. The switch chooses one of the nearest-neighbor nodes to be entangled with, and single-photons are interfered using a 50/50 beam-splitter. Single-photon detectors detect interfered photons leaving electronic spins entangled.}
\centering
\label{fig: SPDschematic}
\end{figure}

The physical unit cell could be very small, on the order of tens of microns, so that the entire lattice may be integrated on a chip, as illustrated in Fig. \ref{fig: SPDschematic}. Each node in the architecture requires an atomic memory and a $1 \times d$ switch, where $d$ is the number of nearest neighbors in the underlying lattice, i.e., the degree of the lattice. Each bond in the lattice requires waveguides, a beam-splitter, and two detectors for the Bell measurement. 

At each time step, a state-selective optical $\pi$-pulse entangles a photonic mode with the electronic spin. Photonic modes coupled to neighboring electronic spins undergo probabilistic linear optic Bell measurement. Successful Bell measurement corresponds to the creation of a bond in the lattice. 

Let us define $n_{\rm att}$ as the total number of entanglement trials per node. Then each bond is attempted $n_{\rm att}/d$ times. If a bond is created before $n_{\rm att}/d$ allocated trials, the node idles on the rest of the attempts for that bond. Each switch only needs to be flipped $d-1$ times during cluster generation, and hence the switching time is negligible compared to the whole process. For example, electro-optic modulators can switch at sub-nanosecond time scales, but the time spent on each bond is on the order of milliseconds as we will see.

The probability of successfully heralding entanglement of two NV centers is $p_0=\eta^2/2$~\cite{2005.PRA.Barrett-Kok.entprot}, where $\eta$ is the efficiency of emitting, transmitting, and detecting the photon entangled with the electronic spin (zero phonon line, ZPL) from the NV excited state. Table~\ref{table:compcoupling} summarizes $p_0$ for three representative types of emitter-photon interfaces: low-efficiency interfaces with $p_0 = 5 \times 10^{-5}$ representative of today's state of the art circular gratings or solid immersion lenses (SILs)~\cite{2013.Nature.Bernien-Hanson.3ment,2015.Nature.Hensen-Hanson.BellTest,2015.NanoLett.Li-Englund.Bullseye}, medium-efficiency interfaces with $p_0 = 2 \times 10^{-4}$  for NV centers coupled to diamond waveguides~\cite{2015.PRX.Mouradian-Englund.DiamWGonSiN}, and high-efficiency $p_0 = 5 \times 10^{-2}$ for nanocavity-coupled NV centers~\cite{2015.NatComm.Li-Englund.NanocavDiam}. For all three coupling mechanisms, we assumed coupling efficiencies that are realistic today~\cite{supp_percatmem}. After $n_{\rm att}/d$ entanglement attempts with a nearest neighbor, the probability of having generated a bond is $p=1-(1-p_0)^{n_{\rm att}/d}$. 

\begin{table} [b]
\caption{\label{Comparison} Bell measurement success probability ($p_0$), Bond trial time ($t_0$) and readout time for three different coupling schemes
} 
\begin{ruledtabular}
\begin{tabular}{l c c c}
Collection&Bullseye or SIL& Waveguide & Cavity\\
\hline

Bond success prob. $p_0$&$5\times10^{-5}$&$2\times10^{-4}$&$5\times10^{-2}$\\

Bond trial time $t_0$\footnote{including state initialization}&5 us&5 us&5 us\\

Readout time&$4$ us&$800$  ns&$100$  ns\\

\end{tabular}
\end{ruledtabular}
\label{table:compcoupling}
\end{table}

\subsection{Percolation thresholds}

\begin{figure}[t]
\includegraphics[width=0.5\textwidth, trim=10 10 10 10]{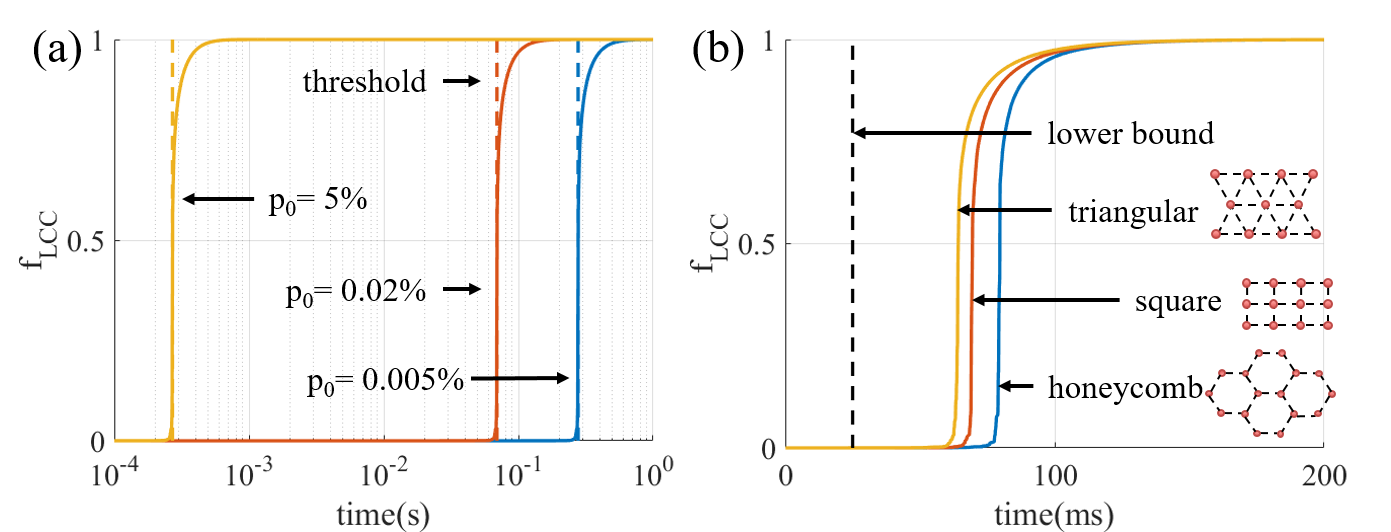}
\caption{Size of the largest connected component vs time for (a) different values of $p_0$, the Bell measurement success probability in one attempt and (b) different underlying lattice geometries. A square lattice is used in (a) and $p_0 = 0.02~\%$ is used for (b).}
\centering
\label{fig:lccvstime}
\end{figure}

We evaluate the growth of clusters using the Newman-Ziff algorithm with $9$ million nodes.  Figure~\ref{fig:lccvstime}(a) plots the fraction of nodes that are in the largest cluster component ($f_{\rm LCC}$), as a function of time from the start of the protocol for the three values of $p_0$, assuming $t_0 = 5~\upmu$s, with an underlying square lattice. Initially, $f_{\rm LCC}$ is O($\log(N)/N$)~\cite{2000.PRE.Bazant.SubCritPerc} where $N$ is the total number of nodes in the lattice. As the bond success probability passes the bond percolation threshold ($p_{c}$), $f_{{\rm LCC}}$ rapidly rises approaching $\Theta(1)$. For a degree $d$ lattice, the bond probability after time $t$ is $p = 1-(1-p_0)^{t/t_0d}$. The time required to obtain a resource for universal quantum computation is $t_c  = t_0d\ln(1-p_c)/\ln(1-p_0)$, which is shown with vertical dashed lines in the figure. The transition becomes sharper as the number of nodes in the lattice ($N$) increases.

In all collection schemes, the bond success probability exceeds the percolation threshold within one second, which is a conservative estimate of the nuclear-spin coherence time in nanostructured, integrated systems based on the experimentally demonstrated coherence time, one minute \cite{bradley2019nuclear.one.minute}. These simulations reveal that an arbitrarily large cluster can be generated even with free space couplings. 

\subsection{Degree of the lattice and imperfection}

It is known that higher degree lattices have a lower percolation threshold. However, $t_{c}$ is nearly the same for the honeycomb ($d=3$), the square ($d=4$) and the triangular ($d=6$) lattices (fig. 3(b)). This is because increasing $d$ lowers the bond percolation threshold, but it also decreases the number of entanglement attempts per bond, $n_{\rm att}/d$; a single broker qubit per NV requires entanglement attempts to proceed serially. Increasing $d$ would in fact substantially lower $t_{c}$ if each site contained multiple broker qubits that could be entangled simultaneously. We do not explore the possibility of multi-NV nodes due to a lack of studies \cite{dolde2013room}. Ion trap systems are presently more mature for this purpose \cite{2014.PRA.Monroe-Kim.ModQC}.

Let's consider the most general case where we can attempt Bell measurements on any pair of NVs at any time step. What is the minimum time, $t^{(LB)}_{c}$, required to obtain a resource for universal quantum computation over all lattice geometries? The bond probability after time $t$ is $p = 1-(1-p_0)^{t/t_0d}$. For percolation, $p \geq p_{c}$ i.e. $t \geq t_0d\cdot {\ln(1-p_{c})}/{\ln(1-p_0)}$. For a degree $d$ lattice, $p_{c} \geq 1/(d-1)$~\cite{1957.MathProcCambPhilSoc.Broadbent-Hammersley.Percolation}, with equality for a degree-$d$ Bethe lattice (infinite tree with fixed degree at each node). This leads to $t_c \geq t_0d\cdot {\ln(1-1/(d-1))}/{\ln(1-p_0)}$. $t_c$ is minimized as $d \rightarrow \infty$ in which case we obtain $t^{(LB)}_c = -{t_0}/{\ln(1-p_0)}$. $t^{(LB)}_{c}$ is plotted as a black dashed line in Fig.~\ref{fig:lccvstime}(b). The lattice corresponding to the lower bound is the infinite-degree Bethe lattice, and such lattices are not a resource for universal quantum computing~\cite{2006.PRL.VanDenNest-Briegel.UnivResMBQC}. Meanwhile, we find that the simple 2D lattices with nearest neighbor connectivity are only a factor $~3$ above this limit and are resources for universal quantum computing. 

\begin{figure}
\includegraphics[width=\columnwidth]{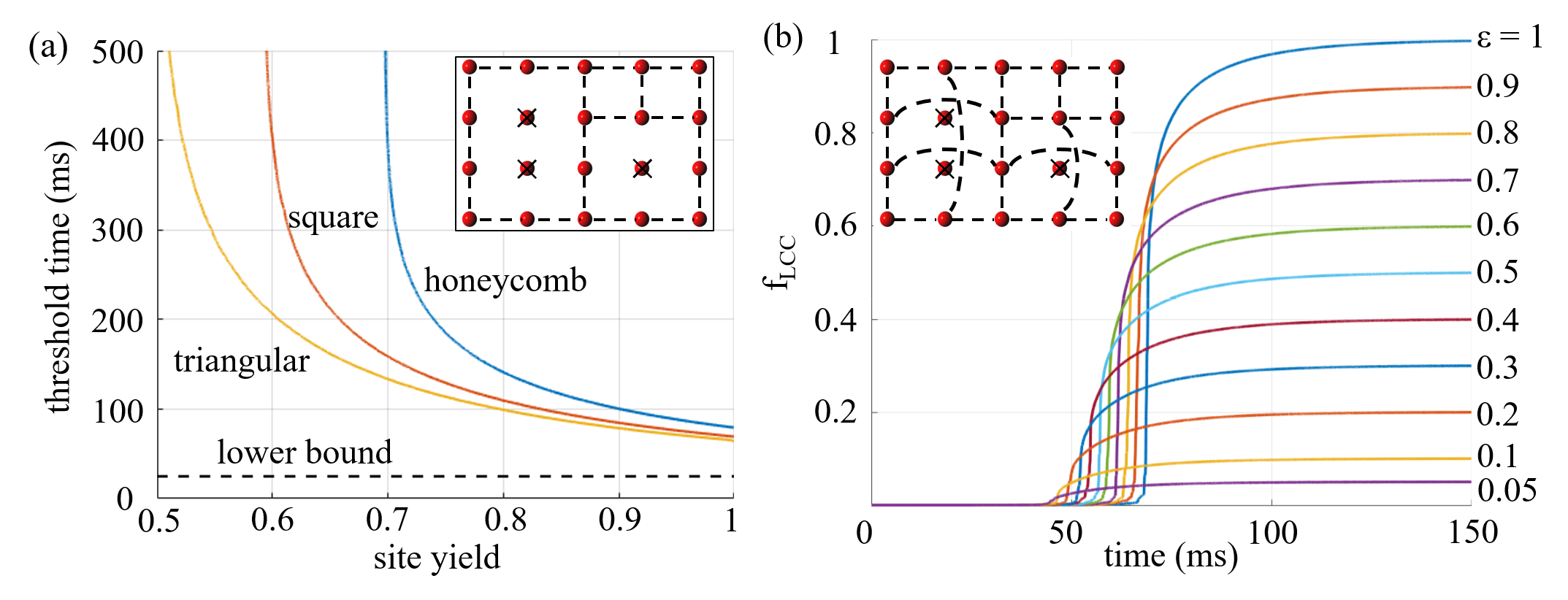}
\caption{(a) The minimum time required to obtain a percolated lattice with sub-unity site-yield.  (b) $f_{\rm LCC}$ as a function of time for different values of $\epsilon$ in the transparent node architecture. The insets show the bonds that can be attempted in a square lattice if the sites marked with crosses are inactive/transparent. $p_0 = 0.02\%$.}
\centering
\label{fig:siteloss}
\end{figure}

Practically, a scalable architecture should be able to tolerate non-functional sites. For example, trapping in a metastable state, a far-detuned transition, failed charge initialization greatly reduced the system performance in the recent state-of-the-art experiment \cite{humphreys2018deterministic}. Even if all faulty nodes and the bonds connected to it are removed, the lattice can retain enough bonds for a percolated lattice (Fig.~\ref{fig:siteloss}(a), inset). The problem maps to site-bond percolation. We define the site-yield $q$ as the fraction of functional nodes. Figure \ref{fig:siteloss}(a) plots the minimum time required to obtain a percolated lattice as a function of $q$, assuming NVs coupled to diamond waveguides ($p_0 = 0.02\%$). In general, a reduced site-yield can be compensated with a larger bond probability which would require a longer time (more attempts). The site percolation threshold, $q_c$, corresponds to the minimum possible site-yield for percolation with all bonds having succeeded ($p = 1$). The triangular ($q_{c} = 0.5$) performs better than the square lattice ($q_{c} \approx 0.593$) that outperforms the honeycomb lattice ($q_{c} \approx 0.697$).

The architecture that we have discussed thus far only allows for nearest neighbor interaction. Adding long range connections shown in the inset of Fig.~\ref{fig:siteloss}(b) can decrease the threshold time and increase tolerance to imperfect site yield. Such an architecture can be implemented by replacing the $1 \times 4$ switch in Fig.~\ref{fig: SPDschematic} with a $5 \times 5$ optical switch, which is depicted in supplementary Fig. S1. Seven Mach-Zehnder interferometers (MZIs) with two phase shifters can implement switching between the set of input and output modes (fig. S1(c))~\cite{supp_percatmem}. The MZI arrays allow a node to be `transparent'. If a node is transparent, photons pass through the node to the next adjacent node for interference (fig. 4(b) inset). By turning multiple adjacent nodes transparent, this bypass enables long range entanglement in a planar architecture.

Interestingly, this transparent node architecture can be used to reduce percolation threshold by randomly turning a fraction $1-\epsilon$ of the nodes transparent. Figure \ref{fig:siteloss}(b) plots $f_{\rm LCC}$ vs. time. As $\epsilon$ decreases, the maximum possible value of $f_{\rm LCC}$ is reduced from one to $\epsilon$ because only a fraction $\epsilon$ nodes are active. However, reduced $\epsilon$ decreases $t_{c}$ because the transparent nodes increase the effective dimensionality of the lattice. Therefore, there is an optimum value of $\epsilon$ that maximizes LCC for a given time. We numerically found a minimum $p_c$ of $0.33$ with transparent nodes, which is achieved when $1/N \ll \epsilon \ll 1$, i.e., $\epsilon\rightarrow 0$ but the number of non-transparent nodes in the lattice is still $\Theta(N)$. Faulty sites can be incorporated into the fraction of transparent nodes as long as the yield exceeds $1/N$.

\begin{figure}[th]
\includegraphics[width=\columnwidth, trim=10 10 10 10]{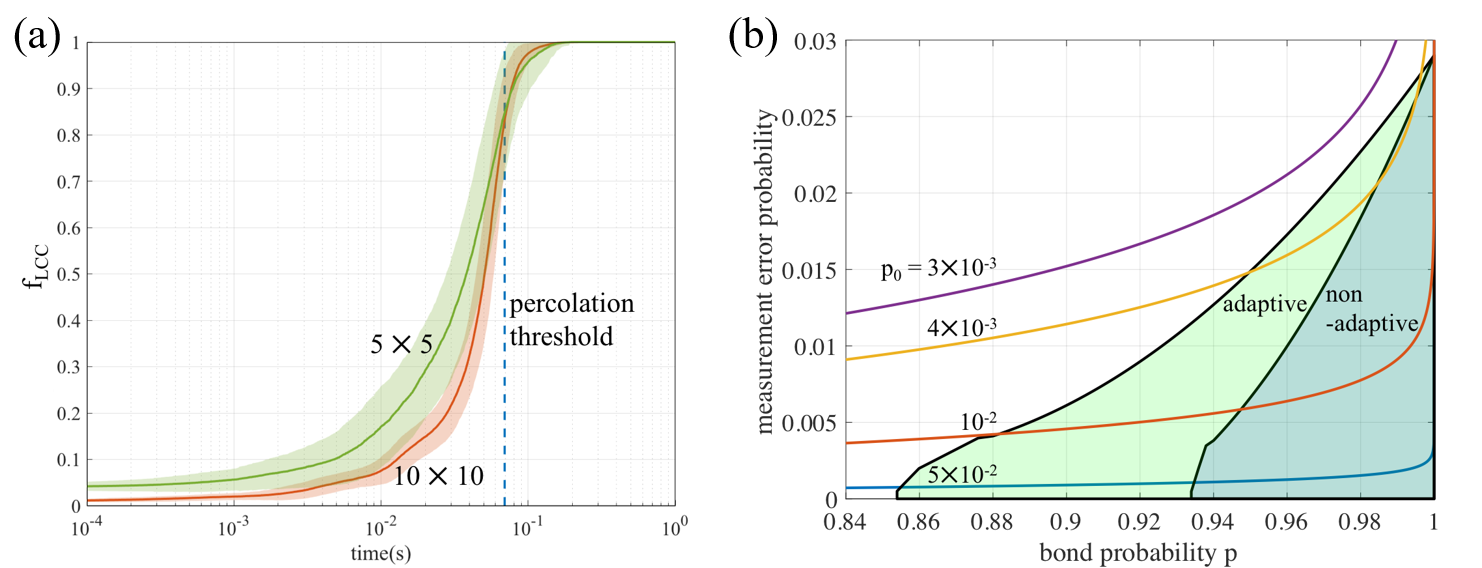}
\caption{(a) Size of LCC as a function of time in a $5 \times 5$ (green) and $10 \times 10$ (red) square grid with $p_0 = 0.02\%$ (corresponding to NVs coupled to diamond waveguides). The lines represent the average size of the LCC and the shaded regions represent one standard deviation. The dashed line is the percolation threshold. (b)The combination of bond probability and measurement error probability achievable for different values of $p_0$. In order to achieve fault tolerance, the curve should have a portion inside the shaded regions which correspond to adaptive and non-adaptive measurement schemes from Ref.~\cite{auger2018fault}}
\centering
\label{Practical}
\end{figure}

\subsection{Small size system}
We used $9$ million qubits (nodes) in the simulations above in order to show the limiting behavior of the cluster creation. However, the same qualitative behavior is also observed in smaller systems. 
In Fig.~\ref{Practical} (a), we plot the mean (line) and the standard deviation (shade) of $f_{LCC}$ as a function of time for a $5 \times 5$ (green) and $10 \times 10$ (red) square lattices ($p_0 = 0.02\%$). In the simulation, we generated 300 lattice instances at each time with a periodic boundary condition and calculated the statistical mean and standard deviation. The transition from small disconnected islands to a large cluster on the order of the lattice size becomes sharper as the size of the system increases, but even for a $5 \times 5$ qubit system, there is a clearly visible transition near the percolation threshold (dashed line). Because of the small system size, there is statistical variation in $f_{LCC}$ and the shaded regions represent one standard deviation. As expected, the relative variation become smaller in larger systems.

\subsection{Fault Tolerance}
It is possible to obtain a regular lattice from the percolated lattices in Fig.~\ref{fig:lccvstime} with a constant overhead by finding crossing paths and using single qubit measurements (renormalization)  \cite{2007.PRL.Kieling-Eisert.PercolationQC, 2015.PRL.Gimeno-Segovia-Rudolph.3GHZtoBallisticQC, 2015.PRA.Zaidi-Rudolph.BallisticLOQC, 2008.NJP.Browne-Short.PercPhaseTrans}. The single qubit measurements used to obtain the regular lattice may bring additional errors, but we can adapt our architecture to the Raussendorf lattice \cite{2006.AnnPhys.Raussendorf-Goyal.FaultTolClus, 2007.NJP.Raussendof-Goyal.TopoFTClus}. The Raussendorf lattice is a 3D lattice with degree-4 (fig. S3 (a)) and a means of translating surface-code error correction into the cluster state model of quantum computation. The Raussendorf lattice can be constructed in a 2+1D architecture \cite{2017.APLPhot.Rudolph.SiPhOptimism} where qubits are arranged in 2D, and an additional dimension is constructed in time (fig. S3 (b)) \cite{2014.PRX.Nemoto-Munro.DiamQCArch,supp_percatmem}. Then, one needs only two layers of 2D lattices to implement a 2+1D lattice because memories are reused after measurement (fig. S3 (c)). This architecture requires a small number of waveguide-crossings, which can be implemented with low loss \cite{han2018high}. Alternatively, they can be absorbed in the optical switches \cite{supp_percatmem}.

Recent results have evaluated the fault-tolerance in the Raussendorf lattice with non-deterministic entangling gates \cite{auger2018fault, herr2018local}. Following \cite{auger2018fault}, fault-tolerance requires the bond probability ($p$) and measurement error to be in the two shaded regions in Fig.~\ref{Practical}(b), which correspond to adaptive and non-adaptive measurement schemes; in the non-adaptive scheme, qubits are measured in the $X$-basis regardless of the failure of bonds; in the adaptive scheme, one of the qubits connected to the missing bond is measured in the $Z$-basis transforming bond-loss to site-loss. The error threshold of each scheme vanishes at $p\sim0.935$ and $p\sim0.855$, respectively. These correspond to the site percolation threshold of the Raussendorf lattice ($q_c\sim0.75$) when lost bonds are transformed to missing sites (see more details in \cite{auger2018fault}).

We assume that the single qubit measurement error probability increases with time as $1-e^{-t/t_{coh}}$, where we use $t_{coh} = 1$ sec. At $t = 0$, the measurement error probability and bond probability ($p$) are both zero. Both of these probabilities increase as we spend more time attempting entanglement generation, resulting in the curves shown in Fig.~\ref{Practical}(b) depending on the entanglement success probability per attempt ($p_0$). Assuming a bond trial time $t_0=5$ $\mu$s (Table~\ref{table:compcoupling}), we find that a value of $p_0 \approx 4 \times 10^{-3}$ is required to meet the fault-tolerance threshold which corresponds to an NV to detector coupling efficiency of $\sim9$ $\%$. 

As an example, we estimated the resource overhead for fault-tolerant factorization of 2000-bit numbers using Shor's algorithm (Fig. 6). Various methods from \cite{fowler2012surface,gidney2018efficient,gidney2019factor,vedral1996quantum,beauregard2002circuit,cuccaro2004new,nielsen_chuang_2010,jones2013low,gidney2018halving,bravyi2005universal,horsman2012surface,fowler2018low,gidney2019flexible} are used for the calculation (See Supplemental Material for more details). Ref.~\cite{gidney2019factor}, in particular, has the least resource overhead even with the additional overhead from spacetime layout and routing. With the assumptions on the error model described in the supplemental \cite{supp_percatmem}, 2000-bit Shor's algorithm requires 2.9 billion (280 million) qubits and 2.2 hours (43 minutes) of computation given a physical error rate $p=10^{-3}~(10^{-5})$. Alternatively, 57 million (3.8 million) physical qubits can run the algorithm in a day (half a day) as a result of space-time trade-off \cite{supp_percatmem}. Essentially, a large number of high purity $T$-gates for modular exponentiation in Shor's algorithm require a formidable amount of physical qubits. On the other hand, a lower bound is set at 12.2 million (1.46 million) physical qubits for 6000 logical qubits (2028 (243) physical qubits per a logical qubit) for distance-25 (8) surface code.

\begin{figure}[h]
\includegraphics[width=\columnwidth]{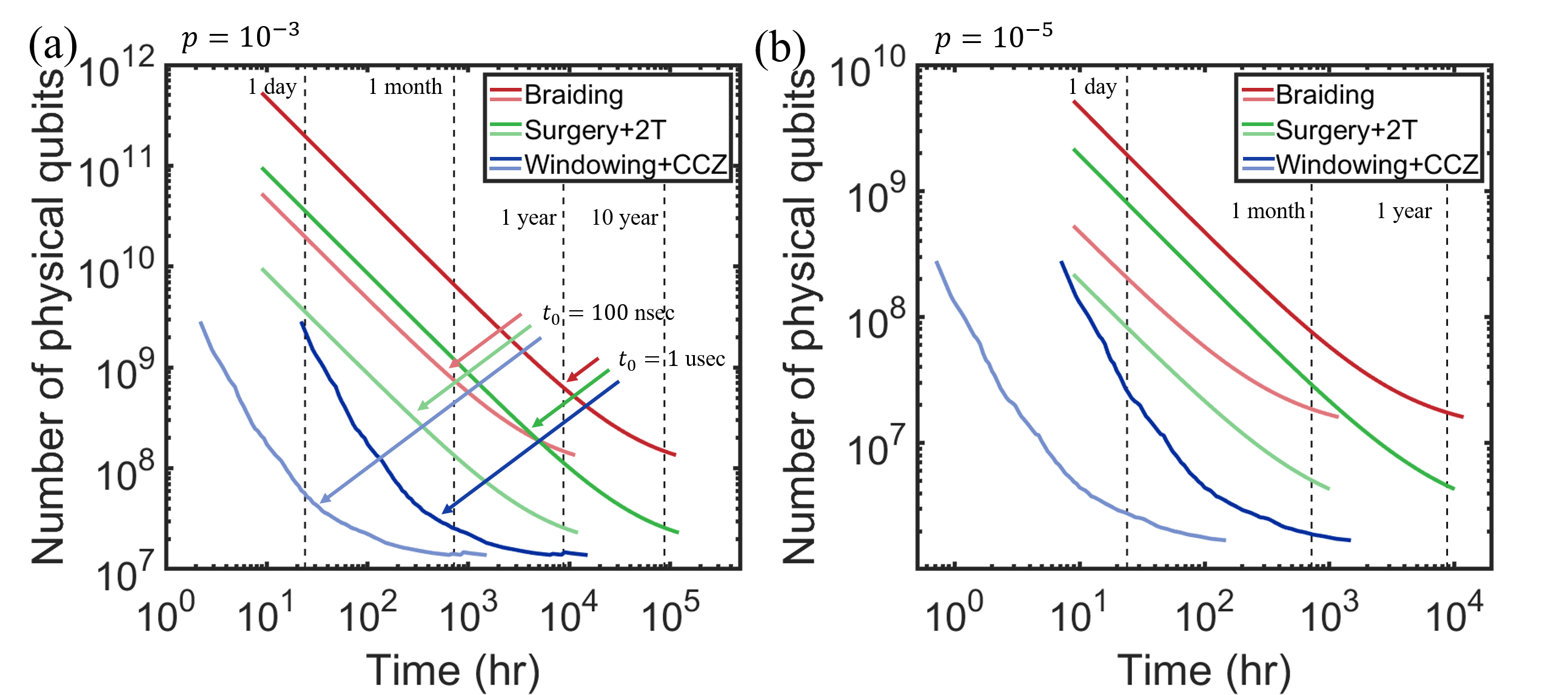}
\caption{Number of physical qubits vs. time for factoring 2000-bit numbers with Shor's algorithm with a physical error rate $p=10^{-3}$ for (a) and $p=10^{-5}$ for (b). Darker lines assume a bond trial time $t_0=1$ $\mu$s, and lighter lines denote $t_0=100$ ns. See Supplemental Information \cite{supp_percatmem} for detailed information and calculation. Results marked with red lines use (double defect) braiding qubits \cite{fowler2012surface} with two-step 15-to-1 distillation for high purity $T$-gate creation. Green lines show results with the lattice surgery qubits \cite{horsman2012surface} and catalyzed-$2T$ factories \cite{gidney2018efficient}. Windowed arithmetic and autoCCZ factories dramatically reduce the resource overhead (blue) \cite{gidney2019factor}. The last result incorporates space-time layout \cite{gidney2019flexible} implying that the improvement is even larger. Results are terminated on the left hand side by either measurement time (red and green) or surface-code cycle (blue) and on the right hand side by logical error rates.
}
\centering
\end{figure}

\section{Discussion}

We assumed here that both bit and phase-flip probabilities are the same. However, our results may be significantly improved using recent work on tailoring the surface code \cite{2017.ArXiv.Tuckett-Flammia.BiasedNoise} if the noise is biased. We consider other important properties of NVs in the supplemental material~\cite{supp_percatmem}. 
For example, NVs can be ionized under strong optical pulses, but their charge state can be recovered by optical repumping. We designed the microwave and optical pulse sequences so that the nuclear spin is not disturbed by failed trials and subsequent spin-state initialization. However, precise control of the microwave and optical transitions presents technical challenges and mark an area of active research~\cite{2016.PRX.Reiserer-Markham.QnetMem, 2014.NatComm.Dolde-Wrachtrup.spinent}. Future work should also refine the error model to explicitly include contributions from the higher order internal dynamics of NVs.

Though there have been huge progress in reducing the resource overhead in fault-tolerant quantum computing, it still requires a large number of physical qubits and long computation time due to slow clock cycles. In this aspect, noisy intermediate scale quantum (NISQ) technology \cite{preskill2018quantum} can be a more near to medium term path for our proposal. Two observations are favorable to the NISQ direction; the proposed system well behaves with only 25 qubits as shown in Fig. 5(a); the LCC size scales linearly with the total number of qubits in the supercritical regime. Thus, the quantum simulation on this architecture quickly reaches thermodynamic behavior \cite{mohammady2018low}.

We emphasize that the architecture is compatible with any quantum memories with enough coherence time. Especially, silicon vacancy centers and other group IV emitters such as GeV, SnV or PbV showed promising emission properties \cite{trusheim2019lead,trusheim2018transform}. For example, silicon vacancy centers have higher Debye-Waller factor of $\sim0.7$, with narrow inhomogeneous distribution of $\sim51$ GHz \cite{schroder2017scalable}. Emission wavelength of emitters can be matched by strain tuning with tuning range upto $\sim 440$ GHz \cite{sohn2018controlling}. The spin coherence time of $10$ ms has been shown at dilution fridge temperature \cite{sukachev2017silicon}.

\section{Data Availability}
All the numerical data presented in this paper is the results of C simulations. The code used to generate this data will be made available to the interested reader upon request.

\begin{acknowledgments}
We would like to thank Simon Devitt for helpful comments on the manuscript. H.C., M.P. and D.E. acknowledge support from the Air Force Office of Scientific Research MURI (FA9550-14-1-0052) and the Army Research Laboratory (ARL) Center for Distributed Quantum Information (CDQI). H.C. and D.E. acknowledge support from the Defense Advanced Research Projects Agency (DARPA) DRINQS (HR001118S0024) and the National Science Foundation (NSF) RAISE TAQS (CHE-1839155) and EFRI ACQUIRE (EFMA-1838911). H.C. was also supported in part by a Samsung Scholarship. S. G. acknowledges support from the Office of Naval Research MURI (N00014-16-C-2069). 

H.C. and M.P. contributed equally to this work.

\end{acknowledgments}

\section{Author Contributions}
H.C., M.P. and D.E. conceived the idea. M.P. derived the analytical limits. H.C. and D.E. analyzed the properties and entanglement protocols of NVs as a physical layer of the architecture. H.C. estimated resource overheads. H.C, M.P. and S.G. developed the simulation codes. D.E. supervised the project.

\section{Competing interests}
The authors declare that there are no competing interests.

\bibliography{PercolationAtomicMemoryPaper}

\end{document}